\documentclass[a4paper,fleqn]{cas-dc}

\usepackage[authoryear]{natbib}

\usepackage[subrefformat=parens]{subcaption}

\usepackage{lineno}
\newcommand*\patchAmsMathEnvironmentForLineno[1]{%
  \expandafter\let\csname old#1\expandafter\endcsname\csname #1\endcsname
  \expandafter\let\csname oldend#1\expandafter\endcsname\csname end#1\endcsname
  \renewenvironment{#1}%
     {\linenomath\csname old#1\endcsname}%
     {\csname oldend#1\endcsname\endlinenomath}}%
\newcommand*\patchBothAmsMathEnvironmentsForLineno[1]{%
  \patchAmsMathEnvironmentForLineno{#1}%
  \patchAmsMathEnvironmentForLineno{#1*}}%
\AtBeginDocument{%
\patchBothAmsMathEnvironmentsForLineno{equation}%
\patchBothAmsMathEnvironmentsForLineno{align}%
\patchBothAmsMathEnvironmentsForLineno{flalign}%
\patchBothAmsMathEnvironmentsForLineno{alignat}%
\patchBothAmsMathEnvironmentsForLineno{gather}%
\patchBothAmsMathEnvironmentsForLineno{multline}%
}

\usepackage{ulem}

\newcommand\Erase{\bgroup\markoverwith{\textcolor{red}{\rule[.5ex]{2pt}{0.4pt}}}\ULon}

\newcommand\EraseTwo{\bgroup\markoverwith{\textcolor{magenta}{\rule[.5ex]{2pt}{0.4pt}}}\ULon}

\usepackage{amsmath}   
\usepackage{mathtools} 
\usepackage{cleveref}  

\usepackage{bm}

\usepackage{algorithm}
\usepackage{algpseudocode}

\ExplSyntaxOn
\keys_set:nn { stm / mktitle } { nologo }
\ExplSyntaxOff

\crefname{equation}{Eq.}{Eqs.}%
\crefname{figure}{Fig.}{Figs.}%
\crefname{table}{Tab.}{Tabs.}
\crefformat{section}{Sec.#2#1#3}

\def\tsc#1{\csdef{#1}{\textsc{\lowercase{#1}}\xspace}}
\tsc{WGM}
\tsc{QE}
\tsc{EP}
\tsc{PMS}
\tsc{BEC}
\tsc{DE}

\newcommand{\realset}{\mathbb{R}}

\newcommand{\diag}{\mathrm{diag}}

\newcommand{\fMMG}{f_{\mathrm{MMG}}}
\newcommand{\vm}{v_{\mathrm{m}}}
\newcommand{\Lpp}{L_{\mathrm{pp}}}

\newcommand{\nP}{n_{\mathrm{P}}}
\newcommand{\deltaf}{\delta_{\mathrm{f}}}
\newcommand{\alphaR}{\alpha_{\mathrm{R}}}

\newcommand{\thetat}{\theta^{\mathrm{t}}}
\newcommand{\thetahat}{\hat{\theta}}
\newcommand{\Thetahat}{\hat{\Theta}}

\newcommand{\markupdraft}[2]{
    \ifthenelse{\equal{#1}{display}}{#2}{}
    \ifthenelse{\equal{#1}{color}}{\color{#2}}{}
}

\begin{document}
\let\WriteBookmarks\relax
\def\floatpagepagefraction{1}
\def\textpagefraction{.001}
\shorttitle{Parameter fine-tuning method for MMG model using real-scale ship data}
\shortauthors{Suyama et~al.}

\title [mode = title]{
    Parameter fine-tuning method for MMG model using real-scale ship data
}                   



\author[1]{Rin Suyama}
    \cormark[1]
    \credit{
        Conceptualization,
        Formal analysis,
        Methodology,
        Software,
        Visualization,
        Writing - Original Draft
    }
    \address[1]{Department of Naval Architecture and Ocean Engineering, Graduate School of Engineering, Osaka University, 2-1 Yamadaoka, Suita, Osaka 565-0971, Japan}
    \ead{suyama_rin@naoe.eng.osaka-u.ac.jp}

\author[2]{Rintaro Matsushita}
    \credit{
        Data Curation,
        Project administration,
        Writing - Review \& Editing
    }
    \address[2]{
        Monohakobi Technology Institute Co., Ltd.,
        Yusen Building, 3-2 Marunouchi, 2-Chome, Chiyoda-ku, Tokyo 100-0005, Japan
    }
    \ead{rintaro_matsushita@monohakobi.com}

\author[2]{Ryo Kakuta}
    \credit{
        Data Curation,
        Project administration,
        Writing - Review \& Editing
    }
    \ead{ryo_kakuta@monohakobi.com}

\author[1]{Kouki Wakita}
    \credit{
        Methodology,
        Writing - Review \& Editing
    }
    \ead{kouki_wakita@naoe.eng.osaka-u.ac.jp}

\author[1]{Atsuo Maki}
    \credit{
        Funding acquisition,
        Project administration,
        Supervision,
        Writing - Review \& Editing
    }
    \cormark[1]
    \ead{maki@naoe.eng.osaka-u.ac.jp}

\cortext[cor1]{Corresponding author}

\begin{abstract}
    In this paper, a fine-tuning method of the parameters in the MMG model for the real-scale ship is proposed.
    In the proposed method, all of the arbitrarily indicated target parameters of the MMG model are tuned simultaneously in the framework of SI using time series data of real-sale ship maneuvering motion data to steadily improve the accuracy of the MMG model.    
    Parameter tuning is formulated as a minimization problem of the deviation of the maneuvering motion simulated with given parameters and the real-scale ship trials, and the global solution is explored using CMA-ES.
    By constraining the exploration ranges to the neighborhood of the previously determined parameter values, the proposed method limits the output in a realistic range.
    The proposed method is applied to the tuning of 12 parameters for a container ship with five different widths of the exploration range.
    The results show that, in all cases, the accuracy of the maneuvering simulation is improved by applying the tuned parameters to the MMG model, and the validity of the proposed parameter fine-tuning method is confirmed.
\end{abstract}




\maketitle

\sloppy

\section{Introduction}

    The development of maneuvering models is essential to the research and the development of autonomous navigation algorithms.
    Autonomous navigation algorithms, studied intensively by institutions in many countries in recent years, should be validated by maneuvering simulators.
    The reliability of the algorithm can be improved by testing it under maneuvering simulators with accurate maneuvering models.

    Research on modern maneuvering models has a long history, with literature dating back to the 1940s, such as \citet{davidson1946}.
    The maneuvering models have been studied in terms of hydrodynamics by \citet{Motora1959En,Inoue1981} and control engineering by \citet{Motora1955,Nomoto1957En}.
    In addition, researches have been conducted on models of the hydrodynamic forces of individual components, such as hull, propeller, and rudder.
    For instance, \citet{Abkowitz1964,Astrom1976} utilized Taylor expansion with respect to state variables to derive a maneuvering model.
    The MMG model \citep{Ogawa1978,Yasukawa2015} was groundbreaking in that it could represent each individual component and their interaction.
    \citet{Fossen2011} formulated a maneuvering model in vector form inspired by mechanical systems, and this formulation is often used in model-based control design.
    In addition, there exist studies on models expressed with polynomials of motion states and actuator states \citep{Miyauchi2023,Dong2023}.
    In contrast to these white box models, some studies have addressed the approximation of maneuvering models using black box models based on machine learning.
    Some examples of the black box models are artificial neural networks (ANN) \citep{Moreira2003,Rajesh2008,Zhang2013,Oskin2013,Luo2016,Woo2018,He2022,Wakita2022,Wang2023}, support vector machine \citep{Luo2009,Wang2020}, and Gaussian process \citep{Ramirez2018,Xue2020}.

    Among various maneuvering models, the MMG model has high accuracy in simulations of maneuvering motion.
    This is explained by the fact that it is formulated based on detailed hydrodynamics.
    Due to its high performance, the MMG model is widely used for the design and validation of autonomous navigation algorithms.
    \citet{Li2013} validated the path following control law based on sliding mode control by numerical simulations following the MMG model.
    \citet{Zhang2017} validated the course keeping control law with nonlinear feedback in a simulation based on the MMG model.
    \citet{Zheng2022} designed a path following control of an unmanned surface vessel by applying the framework of reinforcement learning with the state transition model based on the MMG model.
    There are also examples of simulations based on the MMG model that solve the autonomous berthing/unberthing problems in the framework of off-line trajectory planning \citep{Maki2020,Miyauchi2022traj,Suyama2022}.

    The MMG model contains constant parameters to be identified.
    There have been various methods for the identification of parameters in the MMG model for model-scale ships.
    Classically, captive tests were conducted to identify the parameters \citep{Ogawa1978,Yasukawa2015}.
    Computational fluid dynamics (CFD) is also used for the identification of the parameters in the MMG model \citep{Zhang2019}.
    \citet{Sakamoto2019} identified the values of the parameters in the MMG model based on the captive model tests simulated with CFD.

    So far in most cases, the parameters for the real-scale ship MMG model were identified based on the results of model-scale captive tests.
    One of the differences between the maneuvering motion of model-scale ships and real-scale ships is the difference in Reynolds numbers.
    To correct the effect of the difference in Reynolds numbers, for instance, for limited parameters such as the frictional resistance coefficient for an equivalent flat plate and the wake coefficient in straight moving, there are widely used methods such as the three-dimensional extrapolation method (e.g. \citep{Yasukawa2015}).

    However, at this time, the methodology for the identification of the parameters in the real-scale ship MMG model using captive tests is not established.
    Some parameters, such as the flow straightening coefficient, are sometimes used without correcting for the effect of the difference in Reynolds numbers, although the effects have been noted \citep{Aoki2006}.
    In addition, the effects of the difference in Reynolds numbers are not completely clarified for all parameters in the MMG model.
    If captive tests on real-scale ships were conducted, the parameters could be identified without being affected by the difference in Reynolds numbers, but captive tests on large ships, such as containers, are not possible.

    Another method for identification of the parameters in the real-scale ship MMG model is a heuristic tuning of previously determined values by referring to the maneuvering motion data of the real-scale ship.
    In most cases, this tuning is implemented by experienced engineers using their own heuristic methods based on their senses \citep{Sutulo2014}.
    This heuristic tuning directly refers to real-scale ship maneuvering motion data, so the accuracy of the MMG model with the tuned parameters can be steadily improved.
    However, heuristic tuning of a large number of parameters is not optimal, so an automatic tuning method needs to be established.

    System identification (SI) is applicable for the determination of the values of parameters in the real-scale ship MMG model.
    SI is an identification framework for system parameters using input/output data of the target system.
    Therefore, SI for the parameters in the MMG model of real-scale ships can be implemented only with time series data of the maneuvering motion of the subject real-scale ship.
    SI has been applied to parameter identification of ship maneuvering models.
    Kalman filter is one of the popular methods for SI, and \citet{Astrom1976,Abkowitz1980} applied Kalman filter to the Abkowitz model.
    A combination of state estimation and SI was proposed by \citep{Yoon2003} for parameter identification.
    \citet{Araki2012} validated the SI method using the maneuvering motion data simulated using CFD.
    \citet{Sutulo2014} adopted the genetic algorithm as the exploration method for the optimal values of parameters in SI.
    This study was extended to SI for the maneuvering model with a modular structure \citep{Sutulo2023}.
    \citet{Miyauchi2022} proposed a parameter identification method for the MMG model without the initial guess of the parameters using the SI framework.
    This study was extended to the parameter identification of the Abkowitz model with more than 200 parameters and validated its effectiveness \citep{Miyauchi2023}.
    However, in the method proposed by \citet{Miyauchi2022,Miyauchi2023}, maneuvering motion data with random inputs, which is difficult to obtain in the operation of real-scale ships, were used, and the validation of the method was limited to model-scale ship cases.
    Several studies on SI utilize time series data of real-scale ship maneuvering motion \citep{Abkowitz1980,Kim2018,Meng2022,Kambara2022,Wang2023}, but these studies have focused on certain components of the maneuvering model, linear maneuvering models, or ANN models.
    To the best of the authors' knowledge, the SI method for the whole MMG model using real-scale ship data is not discussed.
    

    In this study, the authors propose a fine-tuning method for all of the arbitrarily indicated target parameters of the MMG model using the framework of SI.
    The proposed method is designed to automatically fine-tune the parameter values previously determined based on hydrodynamics, captive model tests, and CFD to the ones for the real-scale MMG model.
    The proposed method directly refers to the time series data of real-scale ship maneuvering motion and steadily improves the performance of the MMG model with the tuned parameter in terms of the accuracy of the simulated maneuvering motion.
    The parameter tuning is formulated as a constrained minimization problem, and the solution is explored using Covariance Matrix Adaption Evolutionary Computation (CMA-ES) \citep{Hansen2007,Hansen2014}.
    To obtain realistic values for the MMG parameters, the ranges of available values for the target parameters in the fine-tuning problem are constrained to the neighborhood of the previously determined values.
    The proposed parameter fine-tuning method is applied to a container ship with $\Lpp = 83 \mathrm{m}$ and validated.
    The objective of this study is not to clarify the effect of the difference in Reynolds numbers between real-scale ships and model-scale ships, but to establish a practical method for fine-tuning MMG parameters.

    The rest of the manuscript is organized as follows.
    \Cref{sec:notation} describes the notation used in this manuscript.
    The MMG model of the subject ship of this study is detailed in \cref{sec:MMG}.
    The proposed parameter fine-tuning method is described in \cref{sec:method}.
    \Cref{sec:CMA_ES} details the CMA-ES.
    \Cref{sec:results} shows the results of parameter tuning for a container ship.
    The discussion on the proposed method is presented in \cref{sec:discussion}.
    \Cref{sec:conclusion} concludes this manuscript.

\section{Notation}

    \label{sec:notation}

    $\realset$ represents the set of all real numbers.
    $|x|$ represents the absolute value of $x \in \mathbb{R}$.
    The overdot `` $\dot{}$ '' represents the derivative with respect to time $t$.
    $A^{\top}$ represents the transposed array of the array $A$.

\section{MMG model}

    \label{sec:MMG}

\begin{figure}[tb]
        \centering
        \includegraphics[width=0.7\hsize]{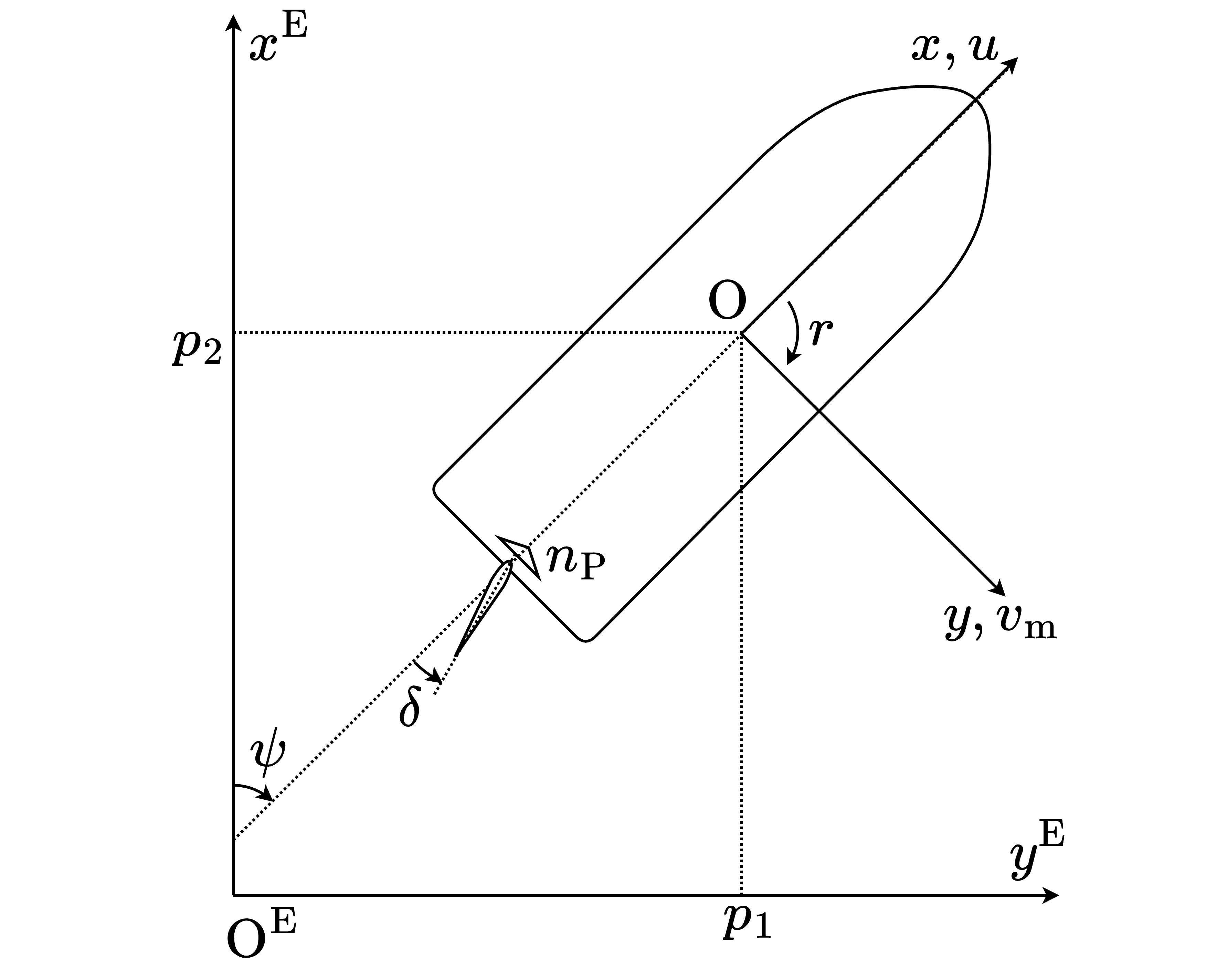}
        \caption{Coordinate systems.}
        \label{fig:coordinate}
    \end{figure}

    First, coordinate systems are defined as \cref{fig:coordinate} shows.
    $\mathrm{O}-xy$ in \cref{fig:coordinate} represents a body-fixed coordinate system with the origin fixed at the midship.
    The ship is assumed to move on an Earth-fixed coordinate system $\mathrm{O}^{\mathrm{E}} - x^{\mathrm{E}} y^{\mathrm{E}}$.
    In this study, the subject ship was a container ship \textit{M. V. SUZAKU}.
    The principal particulars of the subject ship are summarized in \cref{tab:pp}.
    \begin{table}[tb]
        \centering
        \caption{Principal particulars of the subject ship.}
        \begin{tabular}{c|c}
            \hline
            Item & Value \\
            \hline
            $\Lpp ~ \mathrm{[m]}$ & $83.0$  \\
            $B ~ \mathrm{[m]}$ & $13.5$  \\
            $d ~ \mathrm{[m]}$ & $3.8$  \\
            $C_{\mathrm{b}}$ & $0.737$  \\
            $x_{\mathrm{G}} ~ \mathrm{[m]}$ & $0.93$  \\
            $D_{\mathrm{P}} ~ \mathrm{[m]}$ & $2.80$  \\
            $H_{\mathrm{R}} ~ \mathrm{[m]}$ & $3.49$  \\
            $A_{\mathrm{R}} ~ \mathrm{[m^{2}]}$ & $6.282$  \\
            \hline
        \end{tabular}
        \label{tab:pp}
    \end{table}
    The maneuvering motion of the ship is represented with state variables:
    \begin{equation}
        p(t) := (p_{1}(t) ~ p_{2}(t) ~ \psi(t))^{\top}
        \enspace ,
    \end{equation}
    \begin{equation}
        \xi(t) := (u(t) ~ \vm(t) ~ r(t))^{\top}
        \enspace .
    \end{equation}
    The augmented variable:
    \begin{equation}
        \zeta(t) := ( p(t)^{\top} ~ \xi(t)^{\top} )^{\top} \in \realset^{6}
    \end{equation}
    is defined.
    The control input is defined as
    \begin{equation}
        \tau(t) := (\nP(t) ~ \delta(t))^{\top}
        \enspace .
    \end{equation}
    The physical meaning of each component of these variables is summarized in \cref{tab:symbols}.
    \begin{table}[tb]
        \centering
        \caption{Explanation of symbols.}
        \begin{tabular}{c|c|c}
            \hline
            Vector & Elements & Description  \\
            \hline
            \multirow{3}{*}{$p$}
                & $p_{1}$ & Position on the Earth-fixed frame.  \\
                & $p_{2}$ & Position on the Earth-fixed frame.  \\
                & $\psi$ & Heading angle.  \\
            \hline
            \multirow{3}{*}{$\xi$}
                & $u$ & Surge velocity.  \\
                & $v_{\mathrm{m}}$ & Sway velocity at the midship.  \\
                & $r$ & Yaw angular velocity.  \\
            \hline
            \multirow{2}{*}{$\tau$}
                & $n_{\mathrm{P}}$ & Propeller revolution number.  \\
                & $\delta$ & Rudder angle.  \\
            \hline
        \end{tabular}
        \label{tab:symbols}
    \end{table}
    Each component of these variables follows the coordinate systems in \cref{fig:coordinate}.
    In this study, the authors only consider forward motion with forward propeller revolution; $u(t) > 0$ and $\nP(t) > 0$.
    In addition, current and wind are ignored as they are assumed to have little effect on the maneuvering motion in this study.
    In the following, $(t)$, which indicates the dependence of variables on time, is omitted to simplify the description.

    States $p$ and $\xi$ follow the kinematics equation:
    \begin{equation}
        \label{eq:kinematics}
        \dot{p} = R(p) \xi
        \enspace ,
    \end{equation}
    where
    \begin{equation}
        R(p) := 
            \begin{pmatrix}
                \cos \psi & -\sin \psi & 0  \\
                \sin \psi & \cos \psi & 0  \\
                0 & 0 & 1  \\
            \end{pmatrix}
        \enspace .
    \end{equation}
    The equation of motion of ship maneuvering is formulated as
    \begin{equation}
        \label{eq:eq_of_motion}
        \left \{
        \begin{aligned}
            ( m + m_{x} ) \dot{u} - ( m + m_{y} ) \vm r - x_{\mathrm{G}} m r^{2} &= X  \\
            ( m + m_{y} ) \dot{v}_{\mathrm{m}} + ( m + m_{x} ) u r + x_{\mathrm{G}} m \dot{r} &= Y  \\
            ( I_{\mathrm{g}} + x_{\mathrm{G}}^{2} m + J_{zz} ) \dot{r} + x_{\mathrm{G}} m ( \dot{v}_{\mathrm{m}} + u r ) &= N
        \end{aligned}
        \right .
    \end{equation}
    where $m$, $m_{x}$, $m_{y}$, $I_{\mathrm{g}}$, $J_{zz}$, $x_{\mathrm{G}}$ are the displacement, the added mass for surge motion, the added mass for sway motion, the moment of inertia for yaw motion along the vertical line passing the midship, the added moment of inertia for yaw motion, $x$ coordinate of the center of gravity of the ship, respectively.
    $X$, $Y$, and $N$ in \cref{eq:eq_of_motion} represent the forces and moments, excluding centrifugal forces, on the body fixed coordinate system $\mathrm{O}-xy$.
    The MMG model \citep{Yasukawa2015} represents $X$, $Y$, and $N$ as the sum of the effects of the hull force, propeller force, and rudder force as follows:
    \begin{equation}
        \left \{
        \begin{aligned}
            X(\thetat) &=
                X_{\mathrm{H}}
                + X_{\mathrm{P}}
                + X_{\mathrm{R}}  \\
            Y(\thetat) &=
                Y_{\mathrm{H}}
                + Y_{\mathrm{P}}
                + Y_{\mathrm{R}}  \\
            N(\thetat) &=
                N_{\mathrm{H}}
                + N_{\mathrm{P}}
                + N_{\mathrm{R}}  \\
        \end{aligned}
        \right .
    \end{equation}
    where the argument $\thetat$ represents the target parameter to be tuned.
    The target parameter $\thetat$ is detailed in \cref{sec:target_parameters}.
    The MMG model's representation of hull force, propeller force, and rudder force is unique in that it even takes into account the interference and interaction of each component, as detailed below.
    By reshaping \cref{eq:eq_of_motion}, the system for $\xi$ is obtained as
    \begin{equation}
        \label{eq:xi_system}
        \dot{\xi} = \fMMG(\xi, \tau ; \thetat)
        \enspace ,
    \end{equation}
    where
    \begin{equation}
        \label{eq:f_MMG}
        \begin{split}
            &\fMMG(\xi, \tau ; \thetat)  \\
                &\quad :=
                    M^{-1}
                    \begin{pmatrix}
                        X(\thetat) + ( m + m_{y} ) \vm r + x_{\mathrm{G}} m r^{2} \\
                        Y(\thetat) - ( m + m_{x} ) u r  \\
                        N(\thetat) - x_{\mathrm{G}} m u r  \\
                    \end{pmatrix} \enspace ,
        \end{split}
    \end{equation}
    \begin{equation}
        M :=
            \begin{pmatrix}
                m + m_{x} & 0 & 0  \\
                0 & m + m_{y} & x_{\mathrm{G}} m  \\
                0 & x_{\mathrm{G}} m & I_{\mathrm{g}} + x_{\mathrm{G}}^{2} m + J_{zz}  \\
            \end{pmatrix}
        \enspace .
    \end{equation}
    Augmenting \cref{eq:kinematics} and \cref{eq:xi_system}, the system for $\zeta$ is obtained as
    \begin{equation}
        \label{eq:zeta_system}
        \dot{\zeta} = f_{\zeta}(\zeta, \tau ; \thetat)
        \enspace ,
    \end{equation}
    where
    \begin{equation}
        f_{\zeta}(\zeta, \tau ; \thetat) :=
            \big(
                ~ ( R(p) \xi )^{\top}
                ~ ( \fMMG(\xi, \tau ; \thetat) )^{\top}
                ~
            \big)^{\top}
        \enspace .
    \end{equation}

    The following sections detail the hydrodynamic force model for each component.
    The MMG model used in this study is based on the one proposed by \citet{Okuda2023}.
    However, the model for hull force was the widely used polynomial model.

    In this manuscript, characters with the prime symbol represent nondimensionalized quantities.
    The nondimensionalized values of physical quantities with unit $\mathrm{[m]}$, $\mathrm{[kg]}$, $\mathrm{[kg m^{2}]}$, and $\mathrm{[m/s]}$ are calculated by dividing with $\Lpp$, $0.5 \rho \Lpp^{2} d$, $0.5 \rho \Lpp^{4} d$, and $U$, respectively, where $U := \sqrt{u^{2} + \vm^{2}}$ is the ship speed.
    In the following, $\beta := \arctan(-\vm / u)$ represents the drift angle.

    \subsection{Hull force \citep{Okuda2023}}
        The surge force, the sway force, and the yaw moment induced by the interaction between the ship hull and the fluid were modeled as follows: 
        \begin{equation}
            \label{eq:hull_force}
            \left \{
            \begin{aligned}
                X_{\mathrm{H}} &= \frac{1}{2} \rho \Lpp d U^{2} X'_{\mathrm{H}}(\vm', r')  \\
                Y_{\mathrm{H}} &= \frac{1}{2} \rho \Lpp d U^{2} Y'_{\mathrm{H}}(\vm', r')  \\
                N_{\mathrm{H}} &= \frac{1}{2} \rho \Lpp^{2} d U^{2} N'_{\mathrm{H}}(\vm', r')
            \end{aligned}
            \right .
        \end{equation}
        The nondimensionalized forces and moment were modeled as
        \begin{equation}
            \left \{
            \begin{aligned}
                X'_{\mathrm{H}}(\vm', r')
                    &= -R'_{0} + X'_{vv} \vm'^{2} + X'_{vr} \vm' r' + X'_{rr} r'^{2}  \\
                    & \qquad + X'_{vvvr} \vm'^{3} r' + X'_{vvvv} \vm'^{4}  \\
                Y'_{\mathrm{H}}(\vm', r')
                    &= Y'_{v} \vm' + Y'_{r} r'  \\
                    & \qquad + Y'_{vvv} \vm'^{3} + Y'_{vvr} \vm'^{2} r'  \\
                    & \qquad + Y'_{vrr} \vm' r'^{2} + Y'_{rrr} r'^{3}  \\
                N'_{\mathrm{H}}(\vm', r')
                    &= N'_{v} \vm' + N'_{r} r'  \\
                    & \qquad + N'_{vvv} \vm'^{3} + N'_{vvr} \vm'^{2} r'  \\
                    & \qquad + N'_{vrr} \vm' r'^{2} + N'_{rrr} r'^{3}  \\
            \end{aligned}
            \right .
        \end{equation}
        where the coefficients are constant parameters which are called the hydrodynamic derivatives.

    \subsection{Propeller force \citep{Okuda2023}}

        The propeller force for $u > 0$ and $\nP > 0$ was modeled as
        \begin{equation}
            \left \{
                \begin{aligned}
                    X_{\mathrm{P}} &= \frac{1}{2} \rho S_{\mathrm{P}} V_{r}^{2} (1 - t_{\mathrm{P}}) K_{\mathrm{T}}(\phi_{\mathrm{P}})  \\
                    Y_{\mathrm{P}} &= \frac{1}{2} \rho S_{\mathrm{P}} V_{r}^{2} C_{\mathrm{P} Y}(\phi_{\mathrm{P}})  \\
                    N_{\mathrm{P}} &= \frac{1}{2} \rho S_{\mathrm{P}} \Lpp V_{r}^{2} C_{\mathrm{P} N}(\phi_{\mathrm{P}})  \enspace .
                \end{aligned}
            \right .
        \end{equation}
        Here $t_{\mathrm{P}}$ is a constant parameter called the thrust deduction factor.
        $\phi_{\mathrm{P}}$ is defined as
        \begin{equation}
            \phi_{\mathrm{P}} := \frac{180}{\pi} \arctan \bigg( \frac{u_{\mathrm{P}}}{0.7 \pi \nP D_{\mathrm{P}}} \bigg)
        \end{equation}
        where $u_{\mathrm{P}} := (1 - w_{\mathrm{P}}) u$.
        The function $K_{\mathrm{T}}(\phi_{\mathrm{P}})$ was modeled as
        \begin{equation}
            \begin{split}
                K_{\mathrm{T}}(\phi_{\mathrm{P}})
                    = &3.31 \times 10^{-6} \phi_{\mathrm{P}}^{3} - 3.72 \times 10^{-4} \phi_{\mathrm{P}}^{2}  \\
                    &- 2.60 \times 10^{-3} \phi_{\mathrm{P}} + 0.167  \enspace ,
            \end{split}
        \end{equation}
        by fitting measured data in the propeller open test shown in Fig.6 in the original paper \citep{Okuda2023}.
        The function $C_{\mathrm{P} Y}(\phi_{\mathrm{P}})$, $C_{\mathrm{P} N}(\phi_{\mathrm{P}})$ were modeled as
        \begin{equation}
            \begin{split}
                C_{\mathrm{P} Y}(\phi_{\mathrm{P}})
                    = &-2.83 \times 10^{-5} \phi_{\mathrm{P}}^{2} + 6.04 \times 10^{-4} \phi_{\mathrm{P}}  \\
                    &- 1.28 \times 10^{-2}  \enspace ,
            \end{split}
        \end{equation}
        \begin{equation}
            C_{\mathrm{P} N}(\phi_{\mathrm{P}}) =
                \left \{
                    \begin{alignedat}{1}
                        &\begin{split}
                            &-2.48 \times 10^{-4} \phi_{\mathrm{P}} - 1.70 \times 10^{-3}  \\
                            &\qquad \text{for} ~ \phi_{\mathrm{P}} < 20  \enspace ,
                        \end{split}  \\
                        &\begin{split}
                            &-1.86 \times 10^{-4} \phi_{\mathrm{P}} + 4.03 \times 10^{-3}  \\
                            &\qquad \text{for} ~ \phi_{\mathrm{P}} \geq 20  \enspace ,
                        \end{split}
                    \end{alignedat}
                \right .
        \end{equation}
        by fitting measured data shown in Fig.13 in the original paper \citep{Okuda2023}.
        In addition,
        \begin{equation}
            S_{\mathrm{P}} := \frac{\pi D_{\mathrm{P}}^{2}}{4}
        \end{equation}
        is the propeller disk area, and
        \begin{equation}
            V_{r} := \sqrt{ u_{\mathrm{P}}^{2} + (0.7 \pi \nP D_{\mathrm{P}})^{2} }
        \end{equation}
        is the apparent inflow velocity into the propeller.
        Here the wake fraction factor $w_{\mathrm{P}}$ was modeled as
        \begin{equation}
            w_{\mathrm{P}} = w_{\mathrm{P} 0} \exp( C_{w} \beta_{\mathrm{P}}^{2} )
        \end{equation}
        where
        \begin{equation}
            \beta_{\mathrm{P}} := \beta - l_{\mathrm{P}}' r'
        \end{equation}
        is the geometrical inflow angle of the propeller with $l_{\mathrm{P}}' = -0.5$, $w_{\mathrm{P} 0}$ is the value of $w_{\mathrm{P}}$ at $\beta_{\mathrm{P}} = 0$.

    \subsection{Rudder force \citep{Okuda2023}}
        The surge force, sway force, and yaw moment induced by the rudder were modeled as follows:
        \begin{equation}
            \left \{
                \begin{aligned}
                    X_{\mathrm{R}} &= -(1 - t_{\mathrm{R}}) F_{\mathrm{N}} \sin \delta  \\
                    Y_{\mathrm{R}} &= -(1 + a_{\mathrm{H}}) F_{\mathrm{N}} \cos \delta  \\
                    N_{\mathrm{R}} &= -(x_{\mathrm{R}} + a_{\mathrm{H}} x_{\mathrm{H}}) F_{\mathrm{N}} \cos \delta
                \end{aligned}
            \right .
        \end{equation}
        The rudder normal force $F_{\mathrm{N}}$ was model as
        \begin{equation}
            \label{eq:F_N}
            F_{\mathrm{N}} = \frac{1}{2} \rho A_{\mathrm{R}} U_{\mathrm{R}}^{2} \{ f_{\alpha}(\deltaf) \sin\alphaR + C_{\mathrm{l}0}(\deltaf) \}
        \end{equation}
        where $A_{\mathrm{R}}$ is the rudder area,
        \begin{equation}
            U_{\mathrm{R}} := \sqrt{u_{\mathrm{R}}^{2} + v_{\mathrm{R}}^{2}}
        \end{equation}
        is the resultant inflow velocity of the rudder, and
        \begin{equation}
            \alphaR := \delta - \mathtt{atan2}(v_{\mathrm{R}}, u_{\mathrm{R}})
        \end{equation}
        is the effective inflow angle of the rudder.
        The longitudinal inflow velocity component of the rudder $u_{\mathrm{R}}$ was modeled as
        \begin{equation}
            u_{\mathrm{R}} = \max\{u_{\mathrm{R}}^{*}, u_{\mathrm{R}}^{**}\}
        \end{equation}
        where
        \begin{equation}
            \left \{
                \begin{alignedat}{1}
                    u_{\mathrm{R}}^{*}
                        &= u_{\mathrm{P}} \epsilon
                            \bigg\{
                                \eta_{\mathrm{R}} \kappa
                                \bigg(
                                    \sqrt{ 1 + \frac{8 K_{\mathrm{T}}(\phi_{\mathrm{P}})}{\pi J_{\mathrm{P}}^{2}} } - 1
                                \bigg)
                                +1
                            \bigg\}
                        \enspace ,  \\
                    u_{\mathrm{R}}^{**} &= 0.7 \pi \nP D_{\mathrm{P}} u_{\mathrm{R} 0}
                        \enspace .
                \end{alignedat}
            \right .
        \end{equation}
        Here, $\epsilon$ is the ratio of the wake fraction factors at the propeller and rudder positions, $\kappa$ is an experimental constant, and
        \begin{equation}
            \eta_{\mathrm{R}} := \frac{D_{\mathrm{P}}}{H_{\mathrm{R}}}
        \end{equation}
        is the ratio of the diameter of the propeller to the rudder height.
        The lateral inflow velocity component of the rudder $v_{\mathrm{R}}$ was modeled as
        \begin{equation}
            v_{\mathrm{R}} = \gamma_{\mathrm{R}} ( U \sin\beta - l_{\mathrm{R}} r )
        \end{equation}
        where $l_{\mathrm{R}}$ is an experimental constant, and $\gamma_{\mathrm{R}}$ is a constant determined based on
        \begin{equation}
            \gamma_{\mathrm{R}} =
                \left\{
                    ~ \begin{alignedat}{2}
                        & \gamma_{\mathrm{R p}} \quad &&\text{for} ~ \beta_{\mathrm{R}} > 0  \\
                        & \gamma_{\mathrm{R n}} \quad &&\text{for} ~ \beta_{\mathrm{R}} < 0
                    \end{alignedat}
                \right .
        \end{equation}
        with effective inflow angle
        \begin{equation}
            \beta_{\mathrm{R}} := \beta - l'_{\mathrm{R}} r'
            \enspace .
        \end{equation}
        $f_{\alpha}(\cdot)$ and $C_{\mathrm{l}0}(\cdot)$ in \cref{eq:F_N} was modeled as functions of the flap angle $\deltaf$ with
        \begin{equation}
            \label{eq:f_alpha}
            f_{\alpha}(\deltaf) = f_{\alpha 0} + f_{\alpha 2} \deltaf^{2}
        \end{equation}
        \begin{equation}
            \label{eq:C_l_0}
            C_{\mathrm{l}0}(\deltaf) = C_{\mathrm{l}01} \deltaf + C_{\mathrm{l}03} \deltaf^{3}
        \end{equation}
        where $f_{\alpha 0}$, $f_{\alpha 2}$, $C_{\mathrm{l}01}$, and $C_{\mathrm{l}03}$ are constants.
        
        The flap angle $\deltaf$ is determined depending upon the rudder angle $\delta$ in the manipulation system of the subject ship.
        The relationship between $\delta$ and $\deltaf$ is shown in Fig.5 in the original paper \citep{Okuda2023}.


    \subsection{Pre-determined parameter values}
    
        The parameters included in the MMG model described in the previous sections have values determined based on hydrodynamics, captive model tests, and CFD, without using the time series of real-scale ship trials.
        In this manuscript, these values are referred to as \textit{pre-determined} values.
        The pre-determined values of the parameters used in this study are those given in the original paper \citep{Okuda2023}.

        The pre-determined values of added masses and added moment of inertia are shown in \cref{tab:added}.
        \begin{table}[tb]
            \centering
            \caption{Pre-determined values of the added masses and added moment of inertia \citep{Okuda2023}.}
            \begin{tabular}{c|c}
                \hline
                Item & Value \\
                \hline
                $m'_{x}$ & $0.010$  \\
                $m'_{y}$ & $0.168$  \\
                $J_{zz}'$ & $0.010$  \\
                \hline
            \end{tabular}
            \label{tab:added}
        \end{table}
        Moreover, the pre-determined values of the parameters used in the models of hydrodynamic forces and moments are shown in \cref{tab:pre_h,tab:pre_p,tab:pre_r}.
        \begin{table}[tb]
            \centering
            \caption{Pre-determined values of the parameters in the model of $X_{\mathrm{H}}$, $Y_{\mathrm{H}}$, and $N_{\mathrm{H}}$ \citep{Okuda2023}.}
            \begin{tabular}{cc|cc|cc}
                \hline
                Item & Value & Item & Value & Item & Value  \\
                \hline
                $R'_{0}$ & $0.017$ & $Y'_{v}$ & $-0.329$ & $N'_{v}$ & $-0.106$  \\
                $X'_{vv}$ & $0.009$ & $Y'_{r} - m'_{x}$ & $0.090$ & $N'_{r}$ & $-0.057$  \\
                $X'_{vr} + m'_{y}$ & $0.160$ & $Y'_{vvv}$ & $-0.787$ & $N'_{vvv}$ & $-0.037$  \\
                $X'_{rr}$ & $-0.0164$ & $Y'_{vvr}$ & $-0.022$ & $N'_{vvr}$ & $-0.105$  \\
                $X'_{vvvr}$ & $-0.824$ & $Y'_{vrr}$ & $-0.206$ & $N'_{vrr}$ & $0.012$  \\
                $X'_{vvvv}$ & $-0.114$ & $Y'_{rrr}$ & $0.001$ & $N'_{rrr}$ & $-0.008$  \\
                \hline
            \end{tabular}
            \label{tab:pre_h}
        \end{table}
        \begin{table}[tb]
            \centering
            \caption{Pre-determined values of the parameters in the model of $X_{\mathrm{P}}$, $Y_{\mathrm{P}}$, and $N_{\mathrm{P}}$ \citep{Okuda2023}.}
            \begin{tabular}{c|c}
                \hline
                Item & Value  \\
                \hline
                $t_{\mathrm{P}}$ & $0.080$  \\
                $w_{\mathrm{P} 0}$ & $0.422$  \\
                $C_{w}$ & $-2.0$  \\
                \hline
            \end{tabular}
            \label{tab:pre_p}
        \end{table}
        \begin{table}[tb]
            \centering
            \caption{Pre-determined values of the parameters in the model of $X_{\mathrm{R}}$, $Y_{\mathrm{R}}$, and $N_{\mathrm{R}}$ \citep{Okuda2023}.}
            \begin{tabular}{cc|cc}
                \hline
                Item & Value & Item & Value  \\
                \hline
                $t_{\mathrm{R}}$ & $-0.058$ & $\gamma_{\mathrm{R p}}$ & $0.483$  \\
                $a_{\mathrm{H}}$ & $0.158$ & $\gamma_{\mathrm{R n}}$ & $0.172$  \\
                $x_{\mathrm{H}}'$ & $-0.605$ & $f_{\alpha 0}$ & $2.411$  \\
                $\epsilon$ & $1.27$ & $f_{\alpha 2}$ & $-0.381$  \\
                $\kappa$ & $0.5$ & $C_{\mathrm{l}01}$ & $1.164$  \\
                $u_{\mathrm{R} 0}$ & $0.14$ & $C_{\mathrm{l}03}$ & $-0.381$  \\
                $l'_{\mathrm{R}}$ & $-0.888$ & &  \\
                \hline
            \end{tabular}
            \label{tab:pre_r}
        \end{table}

\section{Parameter tuning method}

    \label{sec:method}

    In this section, the target parameter to be tuned, the exploration range for the target parameter, the real-scale ship trials used in this study, and the mathematical formulation of the parameter tuning problem are described.

    \subsection{Target parameter}

        \label{sec:target_parameters}

        In this study, among the parameters included in the MMG model shown in \Cref{sec:MMG}, the following 12 parameters were set as the target parameter $\thetat$ to be tuned.
        \begin{equation}
            \label{eq:target_parameter}
            \thetat :=
                (
                    R'_{0} ~ t_{\mathrm{P}} ~ w_{\mathrm{P} 0} ~ C_{w} ~ t_{\mathrm{R}} ~ a_{\mathrm{H}} ~ x'_{\mathrm{H}} ~ \epsilon ~ \kappa ~ l'_{\mathrm{R}} ~ \gamma_{\mathrm{R p}} ~ \gamma_{\mathrm{R n}}
                )^{\top}
        \end{equation}
        The target parameters were limited to the ones that have a big influence on the simulated forces and moment.
        However, the hydrodynamic derivatives were excluded from the target
        for the simplification of the problem.
        On the other hand, $R'_{0}$ was included in the target of the tuning without applying the extrapolation method.
        The objective for this is to obtain an average value of $R'_{0}$ concerning the variation of ship speed due to maneuvering motion.

        The applicability of the proposed method is not limited to the case with the target parameter \cref{eq:target_parameter}.
        The target parameter can be selected arbitrarily by the users.

    \subsection{Exploration range}
    
        \label{sec:exploring_range}
        
        In the parameter tuning, the exploration ranges for all elements of $\thetat$ were limited.
        In this study, it is assumed that all of the parameters in the MMG model of the model-scale ship of the subject ship have their values which have been determined based on hydrodynamics, captive model tests, and CFD.
        The exploration ranges were set to be the neighborhood of those values.
        By constraining the ranges of available values for the target parameters in the fine-tuning problem to the neighborhood of the pre-determined values, the proposed method limits the output in realistic ranges.

        In the following, the candidate value and the pre-determined value of the target parameter $\thetat \in \realset^{12}$ are described as $\thetahat \in \realset^{12}$ and $\theta^{\mathrm{pre}} \in \realset^{12}$, respectively.
        The exploration range for the element of $\thetahat_{i} ~ (i = 1, \cdots, 12)$ was formulated as
        \begin{equation}
            \label{eq:Thetahat_i}
            \Thetahat_{i} := [ ~ \theta^{\mathrm{pre}}_{i} - a_{\mathrm{r}} |\theta^{\mathrm{pre}}_{i}|, ~ \theta^{\mathrm{pre}}_{i} + a_{\mathrm{r}} |\theta^{\mathrm{pre}}_{i}| ~ ]
        \end{equation}
        with $a_{\mathrm{r}} > 0$.
        The present study considered the hydrodynamics underlying the maneuvering motion to some extent by defining the exploration range with reference to the pre-determined values.
        Further investigation of the exploration range for parameters based on the theoretical background is one of the issues for future research.
        The exploration range of $\thetat \in \realset^{12}$ is described as
        \begin{equation}
            \Thetahat := \prod_{i = 1}^{12} \Thetahat_{i}
            \enspace .
        \end{equation}
        Five cases of parameter fine-tuning with $a_{\mathrm{r}} = 0.2, 0.3, 0.4, 0.5, 0.6$ were conducted for the subject ship.

    \subsection{Real-scale ship trials data}
    
\begin{figure*}[tb]
            \centering
            \includegraphics[width=1.0\hsize]{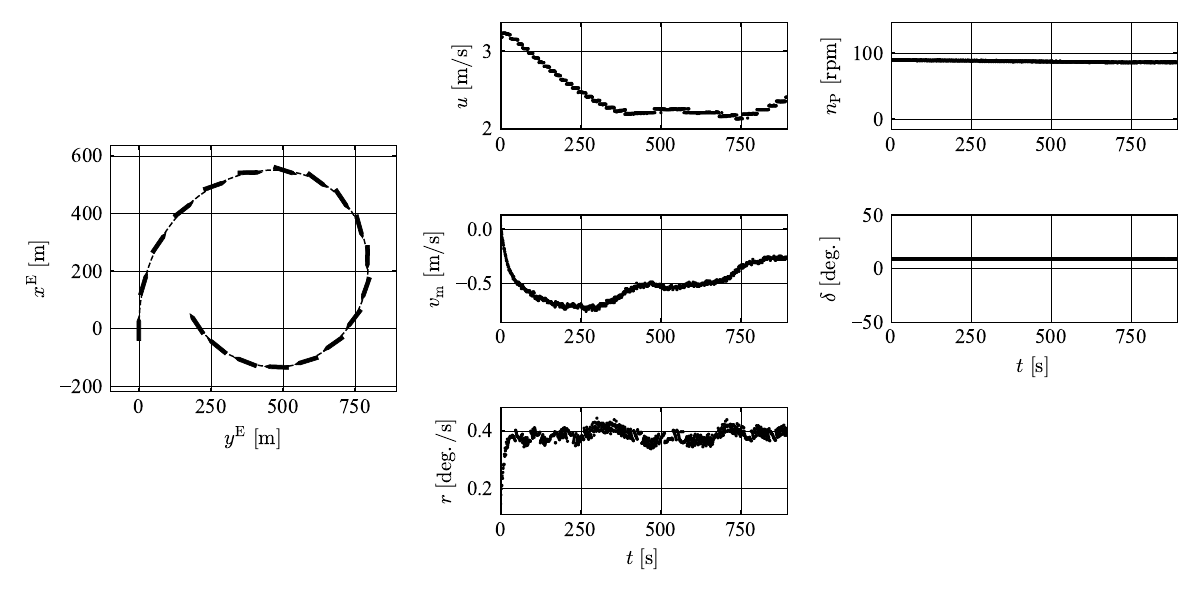}
            \caption{An example of time series data of real-scale ship trial prepared in this study: turning test with $\delta = 10 ~ \mathrm{deg.}$.}
            \label{fig:ts_data_example}
        \end{figure*}

        In this study, the time series of real-scale ship trials were utilized.
        One time series was treated as matrix $D \in \realset^{8 \times T}$ with
        \begin{equation}
            \begin{split}
                ( ~ D_{1 i} ~ \cdots ~ D_{8 i} ~ )^{\top} := &( ~ {p^{i}}^{\top} ~ {\xi^{i}}^{\top} ~ {\tau^{i}}^{\top} ~ )^{\top}  \enspace ,  \\
                &i = 1, \cdots, T  \enspace .
            \end{split}
        \end{equation}
        $T$ is the number of time steps included in the time series data with the time step size: $\Delta t = 1.0 ~ \mathrm{s}$.
        Elements ${p^{i}}$, $\xi^{i}$, and $\tau^{i}$ represent $p$, $\xi$, and $\tau$ at the time step $i$, respectively.
        In this study, eight turning tests of the subject ship with $\delta = \pm 10, ~ \pm 20, ~ \pm 35, ~ \pm 40 ~ \mathrm{deg.}$ were prepared and are described as $D^{\pm 10}, ~ D^{\pm 20}, ~ D^{\pm 35}, ~ D^{\pm 40}$, respectively.
        \Cref{fig:ts_data_example} exemplifies $D^{+10}$.
        
        These time series data were utilized both for the parameter tuning and the test of the tuned parameter.
        In the parameter tuning phase, the parameter that minimizes the deviation between numerically simulated time series and the real-scale ship data was explored.
        In the test phase, the tuned parameter was applied to the simulation of real-scale ship data which was not utilized in the tuning phase and evaluated.
        The sets of time series data for the parameter tuning and the performance test are defined as
        \begin{equation}
            \mathcal{D}^{\mathrm{tune}} := \{ D^{+10}, ~ D^{-20}, ~ D^{+35}, ~ D^{-40} \}
        \end{equation}
        and
        \begin{equation}
            \mathcal{D}^{\mathrm{test}} := \{ D^{-10}, ~ D^{+20}, ~ D^{-35}, ~ D^{+40} \}
            \enspace ,
        \end{equation}
        respectively.

    \subsection{Problem formulation}

        In this study, the performance of the maneuvering model with the given parameter $\thetahat \in \Thetahat$ was defined based on the comparison of the simulated time series of maneuvering motion and the real-scale ship trial data.
        Here, the deviation of the simulated time series following \cref{eq:zeta_system} with the MMG model and the time series of real-scale ship trial was computed, and the candidate parameter $\thetahat$ which minimizes this deviation was output as the optimal value in the parameter tuning problem.
        This minimization problem is summarized as follows.
        \begin{equation}
            \label{eq:minimization_problem}
            \begin{aligned}
                &\underset{ \thetahat \in \Thetahat }{ \text{minimize} }
                    \quad J(\thetahat ; \mathcal{D}^{\mathrm{tune}}) :=
                        \sum_{D \in \mathcal{D}^{\mathrm{tune}}}
                            \sum_{i = 2}^{T} \tilde{p}^{i \top} Q \tilde{p}^{i}  \\
                &\text{s.t.} ~~~
                    \begin{aligned}
                        &\left \{
                            \begin{aligned}
                                \tilde{p}^{i} &:= \hat{p}^{i} - p^{i}  \\
                                \hat{p}^{i} &:= ( \hat{\zeta}^{i}_{1} ~ \hat{\zeta}^{i}_{2} ~ \hat{\zeta}^{i}_{3} )^{\top}  \\
                                \hat{\zeta}^{i} &= \hat{\zeta}^{i-1} + f_{\zeta}(\hat{\zeta}^{i-1}, \tau^{i-1}; \thetahat) \Delta t  \\
                                \hat{\zeta}^{1} &= \zeta^{1}
                            \end{aligned}
                        \right.  \\
                    \end{aligned}  \\
                & \quad \text{for} ~ i = 2, \cdots, T
            \end{aligned}
        \end{equation}
        Here $Q \in \realset^{3 \times 3}$ is a weight matrix for deviation $\tilde{p}$.

        The solution of the minimization problem \cref{eq:minimization_problem} was explored using CMA-ES.
        This exploration algorithm is detailed in \cref{sec:CMA_ES}.
        In the following, the output of CMA-ES for the problem \cref{eq:minimization_problem} is described as $\theta^{*}$.

\section{Covariance Matrix Adaption Evolutionary Computation (CMA-ES)}

    \label{sec:CMA_ES}
    
    CMA-ES \citep{Hansen2007,Hansen2014} is a numerical solver for optimization problems and is known to have high efficiency in the exploration of the optimal solution.
    In the field of ship navigation, CMA-ES has been applied, for instance, to the optimal control problem for autonomous berthing \citep{Maki2020} and the parameter exploration problem for the MMG model based on SI \citep{Miyauchi2022}.
    In general, evolutionary computation does not need the gradient of the objective function.
    Therefore, users do not have to consider the differentiability of the designed objective function.
    Although the minimization problem treated in this study includes the computation of maneuvering motion based on the nonlinear and uncontinuous MMG model, the exploration of the optimal solution with CMA-ES is not inappropriate, as the problem is similar to the one in the paper by \citet{Miyauchi2022} where CMA-ES succeeded in the exploration.
    In the implementation of CMA-ES in this study, the box constraint was handled based on the method proposed by \citet{Sakamoto2017}.

    The algorithm of CMA-ES is summarized as follows.
    The values of the objective function for candidate solutions which are generated following the normal distribution defined with a given mean and a given covariance matrix are iteratively calculated.
    At each step, the algorithm updates the mean and the covariance matrix of the distribution based on the values of the objective function.
    This iterative update is continued until the convergence of the distribution, and the resultant value of the mean is output as the optimal solution.
    In this research, CMA-ES with the restart strategy \citep{Auger2005} was applied.

\section{Results}

    \label{sec:results}

    \subsection{Verification of MMG model with pre-determined parameters}

\begin{figure}[tb]
            \centering
            \includegraphics[width=1.0\hsize]{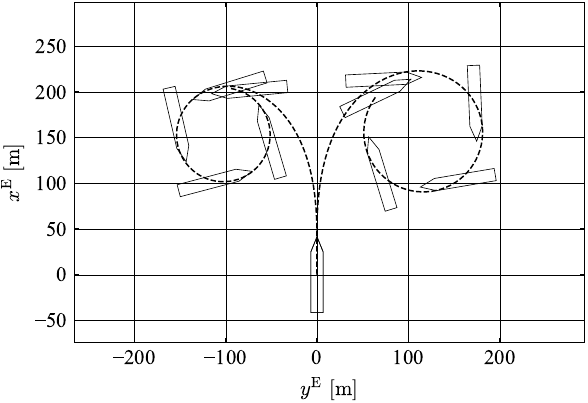}
            \caption{
                The trajectory of turning motion simulated based on the MMG model with pre-determined values ($\delta = \pm 35 ~ \mathrm{deg.}$).
            }
            \label{fig:mmg_verification}
        \end{figure}

        For the verification of the MMG model, the simulation of maneuvering motion was conducted with pre-determined parameters shown in \cref{tab:added,tab:pre_h,tab:pre_p,tab:pre_r}.
        This simulation was calculated by applying Euler method to \cref{eq:zeta_system} with time step size: $\Delta t = 1.0 ~ \mathrm{s}$.
        The initial state was $p(0) = (0 ~ 0 ~ 0)^{\top}$, $u(0) = 6 ~ \mathrm{knots} = 3.086 ~ \mathrm{m/s}$, $\vm(0) = r(0) = 0$.
        The control input was fixed as $\nP = 106 ~ \mathrm{rpm}$, $\delta = \pm 35 ~ \mathrm{deg.}$.
        These conditions were equivalent to the ones set in the simulation shown in Fig.17 (Section 5.3.1) in the original paper \citep{Okuda2023}.
        The simulated time series is shown in \cref{fig:mmg_verification}.
        \Cref{fig:mmg_verification} shows that the simulated turning motion matches the one shown in the original paper \citep{Okuda2023} in that the diameter of the right turning circle is larger than that of the left turning circle.
        This result verifies the MMG model set in this study.

    \subsection{Parameter tuning}

        The weight matrix was set as $Q = \diag(\Lpp, ~ \Lpp, ~ 0.25 \pi)$.
        The initial population size was set $\lambda = 12$, and the population size was doubled at the restart of exploration with the upper limit $\lambda \leq \bar{\lambda} = 128$.

\begin{figure}[tb]
            \centering
            \includegraphics[width=1.0\hsize]{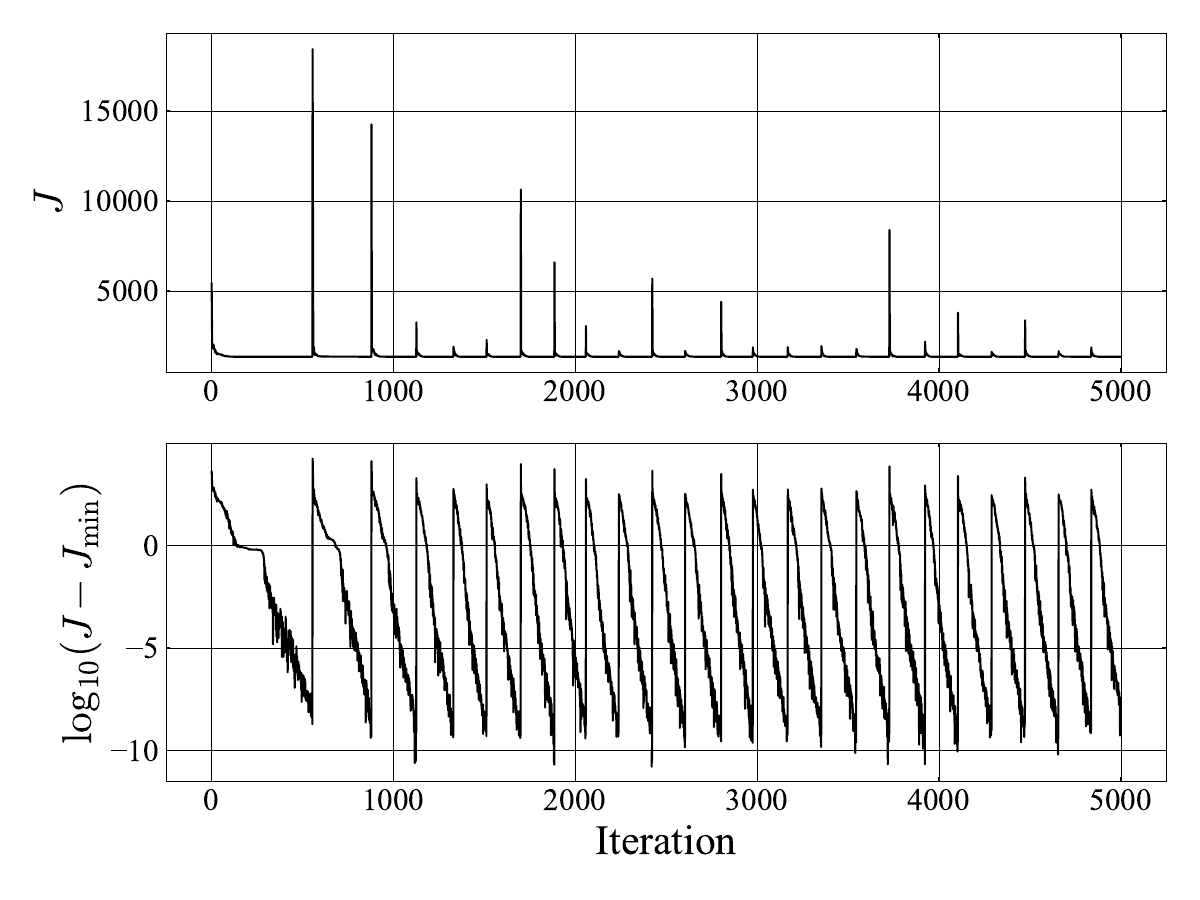}
            \caption{Exploring process of CMA-ES ($a_{\mathrm{r}} = 0.2$).}
            \label{fig:cmaes_process}
        \end{figure}

        First, the process of the exploration by CMA-ES is explained.
        The case of exploration with $a_{\mathrm{r}} = 0.2$ is shown in \cref{fig:cmaes_process}.
        The horizontal axis of \cref{fig:cmaes_process} shows the number of iterations,
        The vertical axis of the upper figure of \cref{fig:cmaes_process} shows the smallest value of the objective function calculated for each of the candidate solutions at the iteration.
        The vertical axis of the lower figure of \cref{fig:cmaes_process} shows the value plotted in the upper figure minus the smallest value of the objective function recorded in all iterations.
        The iteration at which the value of $J$ increases abruptly represents the timing when the convergence of the distribution is confirmed and the exploration restarts with randomly initialized candidate solutions.
        From \cref{fig:cmaes_process}, it can be confirmed that the value of $J$ has repeatedly converged to the same value.
        Not only in the case of $a_{\mathrm{r}} = 0.2$, but also in all the cases conducted in this study, it was confirmed that the optimal solution was explored with a process similar to \cref{fig:cmaes_process}.
        Since multiple restarts were observed in a single trial of the algorithm, and convergence to the same value was observed at each restart, the exploration algorithm for each case was run only once in this study.

\begin{figure}[tb]
            \centering
            \includegraphics[width=1.0\hsize]{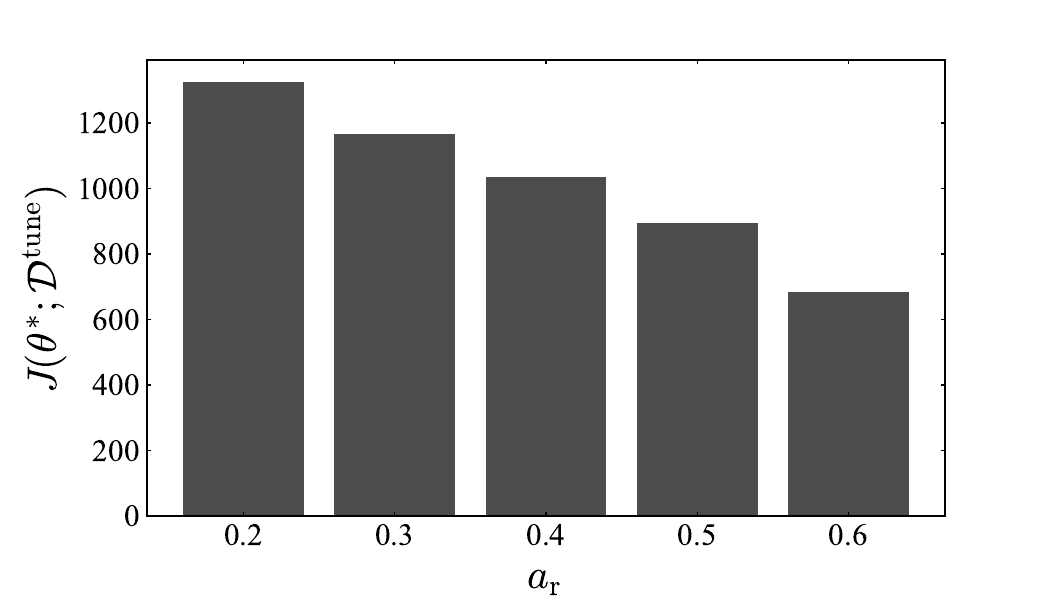}
            \caption{Values of the objective function on the tuned parameter for tuning data $\mathcal{D}^{\mathrm{tune}}$.}
            \label{fig:J_tune_bar}
        \end{figure}

\begin{figure}[tb]
            \centering
            \includegraphics[width=1.0\hsize]{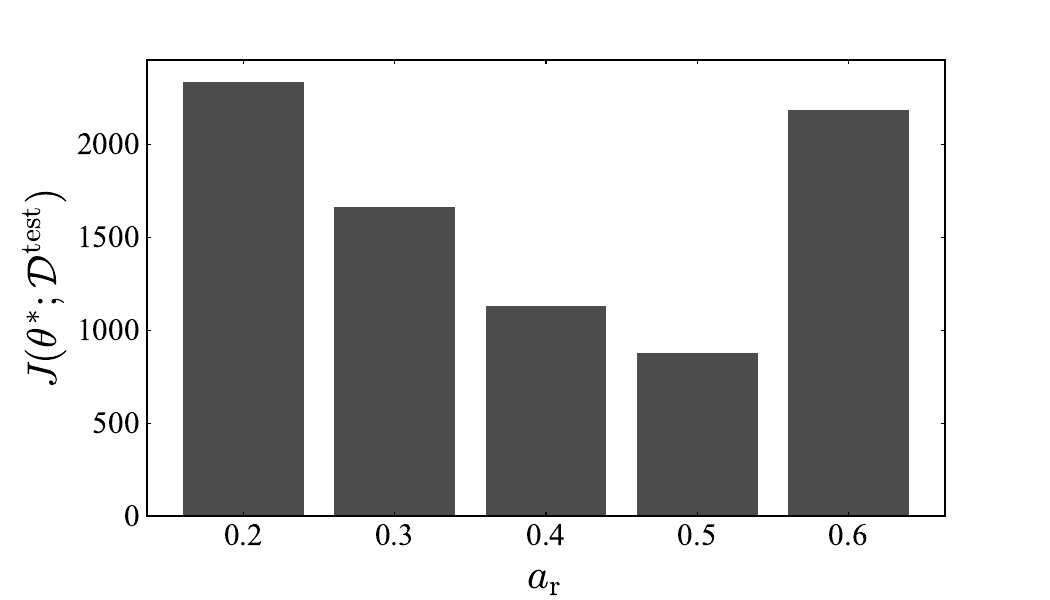}
            \caption{Values of the objective function on the tuned parameter for test data $\mathcal{D}^{\mathrm{test}}$.}
            \label{fig:J_test_bar}
        \end{figure}

        Second, the performance of the tuned parameter is shown.
        For each $\theta^{*}$ explored in each $a_{\mathrm{r}}$, the values of the objective function $ J(\cdot)$ were computed for the data set $\mathcal{D}^{\mathrm{tune}}$ and $\mathcal{D}^{\mathrm{test}}$.
        These values are shown in \cref{fig:J_tune_bar,fig:J_test_bar}.
        It can be confirmed that the value of $J(\theta^{*} ; \mathcal{D}^{\mathrm{tune}})$ decreases as the value of $a_{\mathrm{r}}$ increases.
        This means that the accuracy of the simulated maneuvering motion for the tuning data was improved by expanding the exploration range of the target parameter.
        The value of $J(\theta^{*} ; \mathcal{D}^{\mathrm{test}})$ has the minimum and the maximum at $a_{\mathrm{r}} = 0.5$ and $a_{\mathrm{r}} = 0.6$, respectively.
        In the range from $a_{\mathrm{r}} = 0.2$ to $a_{\mathrm{r}} = 0.5$, the wider the exploration range, the more accurate the MMG model with the tuned parameter applied.
        However, in the case $a_{\mathrm{r}} = 0.6$, while showing the best accuracy for tuning data, the value of $J(\theta^{*} ; \mathcal{D}^{\mathrm{test}})$ is larger than the other cases.
        In the case $a_{\mathrm{r}} = 0.6$, it is considered that the exploration of the parameter that can simulate the maneuvering motion more consistent with the tuning data was performed by expanding the exploration range, resulting in overfitting biased toward the tuning data.

\begin{figure*}[tb]
            \centering
            \includegraphics[width=1.0\hsize]{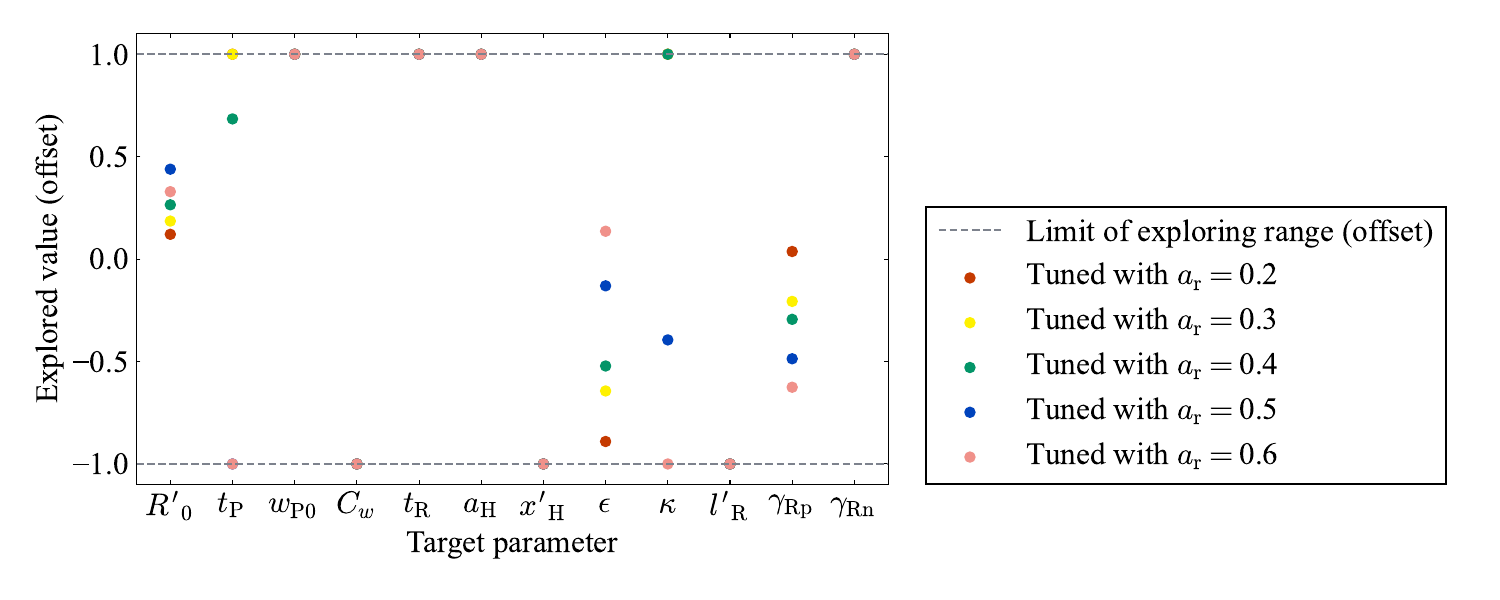}
            \caption{Distribution of the explored values of target parameters in their exploration ranges.}
            \label{fig:explored_values_plot}
        \end{figure*}

        The output values of the target parameters in the exploration with $a_{\mathrm{r}} = 0.2 \sim 0.6$ are shown in \cref{tab:explored_values}.
        \begin{table*}[tb]
            \centering
            \caption{Explored values of target parameters.}
                \begin{tabular}{c|cccccccccccc}
                    \hline
                    $a_{\mathrm{r}}$ & $R'_{0}$ & $t_{\mathrm{P}}$ & $w_{\mathrm{P} 0}$ & $C_{w}$ & $t_{\mathrm{R}}$ & $a_{\mathrm{H}}$ & $x'_{\mathrm{H}}$ & $\epsilon$ & $\kappa$ & $l'_{\mathrm{R}}$ & $\gamma_{\mathrm{R p}}$ & $\gamma_{\mathrm{R n}}$  \\
                    \hline
                    $0.2$ & $0.0174$ & $0.0960$ & $0.5064$ & $-2.4000$ & $-0.0464$ & $0.1896$ & $-0.7260$ & $1.0438$ & $0.6000$ & $-1.0656$ & $0.4866$ & $0.2064$  \\
                    $0.3$ & $0.0179$ & $0.1040$ & $0.5486$ & $-2.6000$ & $-0.0406$ & $0.2054$ & $-0.7865$ & $1.0248$ & $0.6500$ & $-1.1544$ & $0.4531$ & $0.2236$  \\
                    $0.4$ & $0.0188$ & $0.1019$ & $0.5908$ & $-2.8000$ & $-0.0348$ & $0.2212$ & $-0.8470$ & $1.0049$ & $0.7000$ & $-1.2432$ & $0.4262$ & $0.2408$  \\
                    $0.5$ & $0.0207$ & $0.0400$ & $0.6330$ & $-3.0000$ & $-0.0290$ & $0.2370$ & $-0.9075$ & $1.1872$ & $0.4014$ & $-1.3320$ & $0.3655$ & $0.2580$  \\
                    $0.6$ & $0.0204$ & $0.0320$ & $0.6752$ & $-3.2000$ & $-0.0232$ & $0.2528$ & $-0.9680$ & $1.3736$ & $0.2000$ & $-1.4208$ & $0.3017$ & $0.2752$  \\
                    \hline
                \end{tabular}
            \label{tab:explored_values}
        \end{table*}
        In addition, the distribution of each element of the output parameter in its exploration range is shown in \cref{fig:explored_values_plot}.
        In \Cref{fig:explored_values_plot}, the exploration range $\Thetahat_{i}$ is offset to $[-1, 1]$, and the values of each parameter are also offset.
        From \cref{fig:explored_values_plot}, it is observed that, in all cases, $w_{\mathrm{P} 0}$, $C_{w}$, $t_{\mathrm{R}}$, $a_{\mathrm{H}}$, $l'_{\mathrm{R}}$ and $\gamma_{\mathrm{R n}}$ converged to boundary values on the same side of the exploration range, respectively.

\begin{figure*}[tb]
            \centering
            \includegraphics[width=1.0\hsize]{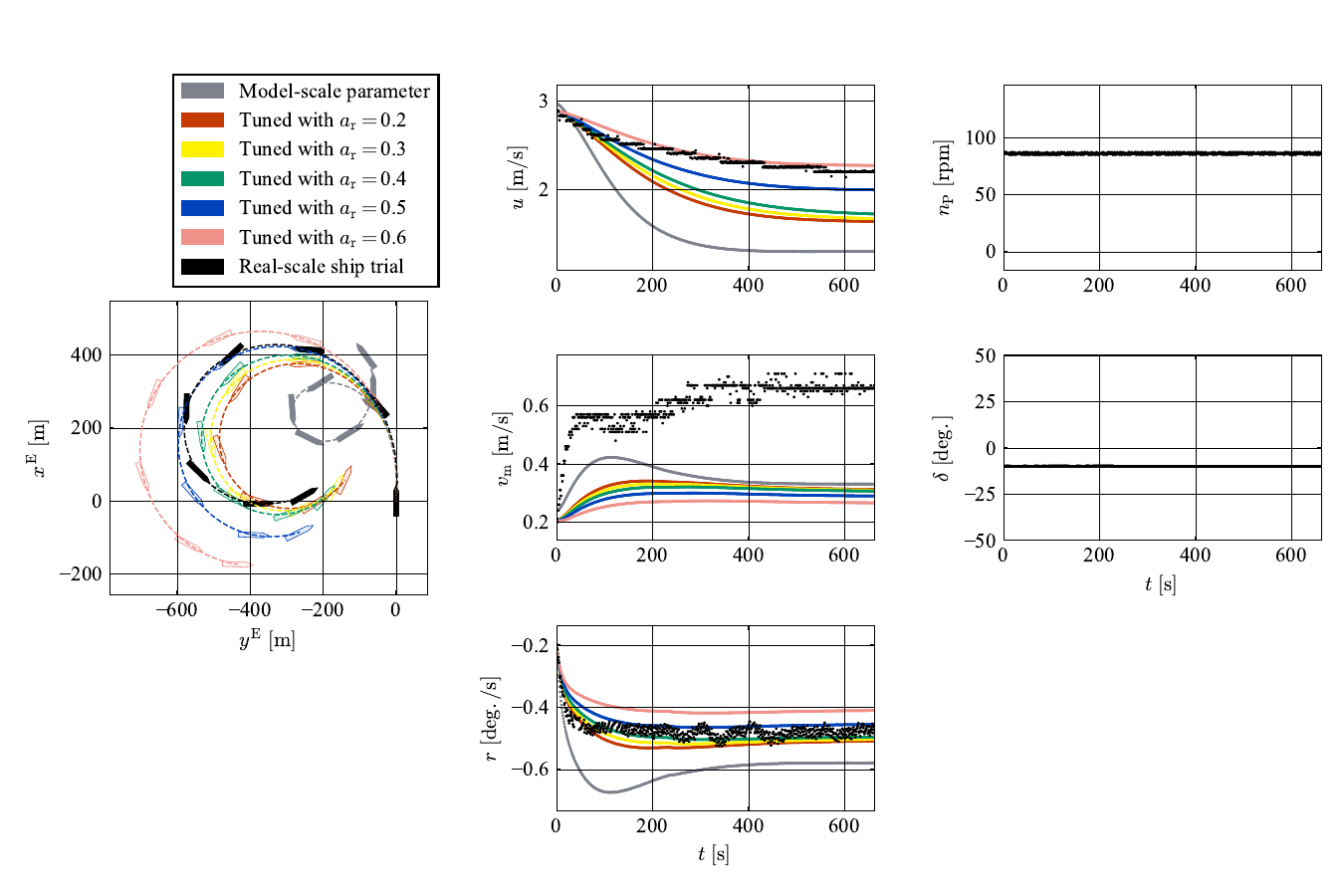}
            \caption{Comparison of seven time series data of turning test ($\delta = -10 ~ \mathrm{deg.}$) which is not utilized in parameter tuning: real-scale ship trial $D^{-10}$, simulation by the MMG model with untuned parameters, and simulation by the MMG model with tuned parameters ($a_{\mathrm{r}} = 0.2 \sim 0.6$).}
            \label{fig:compari_m10}
        \end{figure*}

\begin{figure*}[tb]
            \centering
            \includegraphics[width=1.0\hsize]{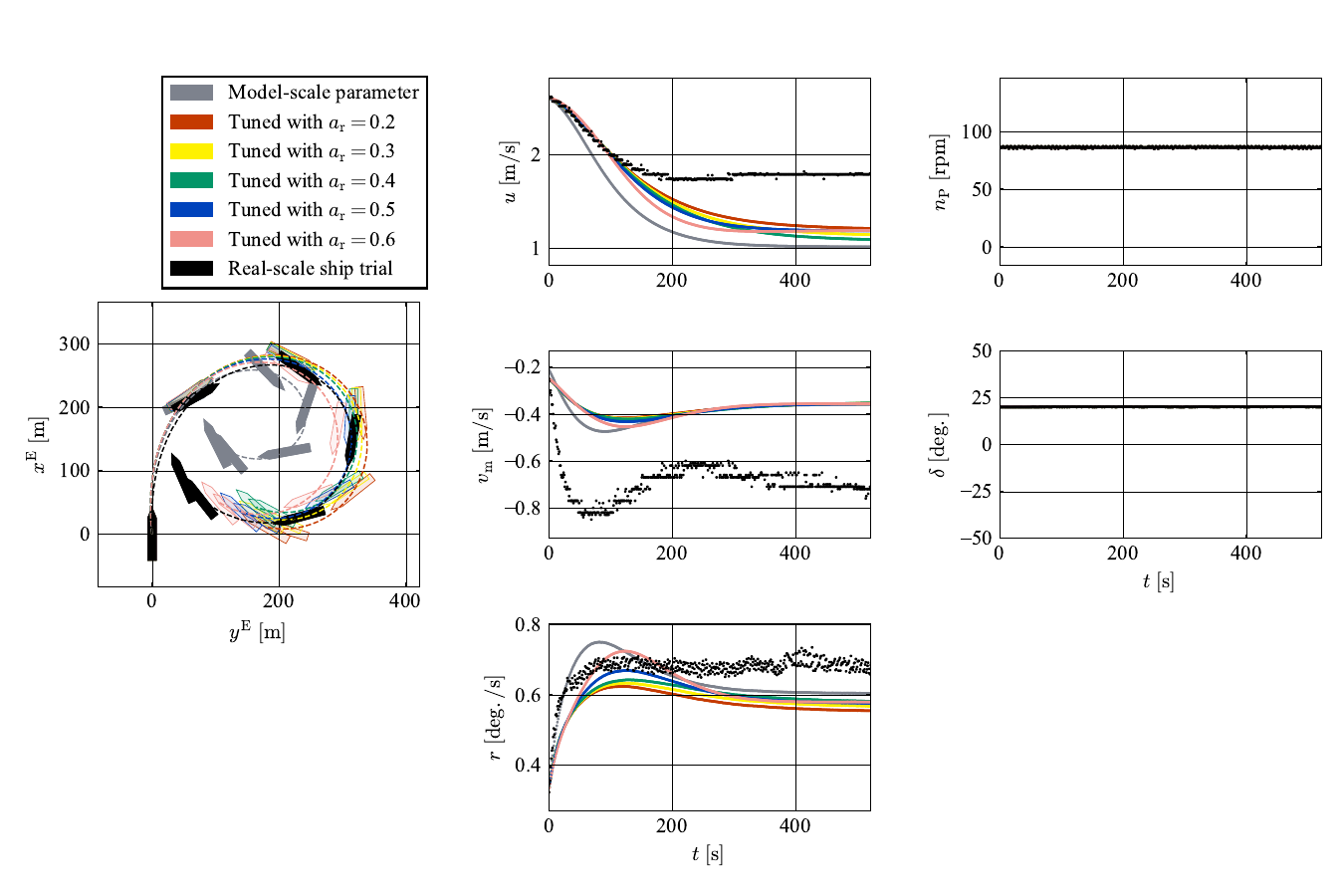}
            \caption{Comparison of seven time series data of turning test ($\delta = +20 ~ \mathrm{deg.}$) which is not utilized in parameter tuning: real-scale ship trial $D^{+20}$, simulation by the MMG model with untuned parameters, and simulation by the MMG model with tuned parameters ($a_{\mathrm{r}} = 0.2 \sim 0.6$).}
            \label{fig:compari_p20}
        \end{figure*}

\begin{figure*}[tb]
            \centering
            \includegraphics[width=1.0\hsize]{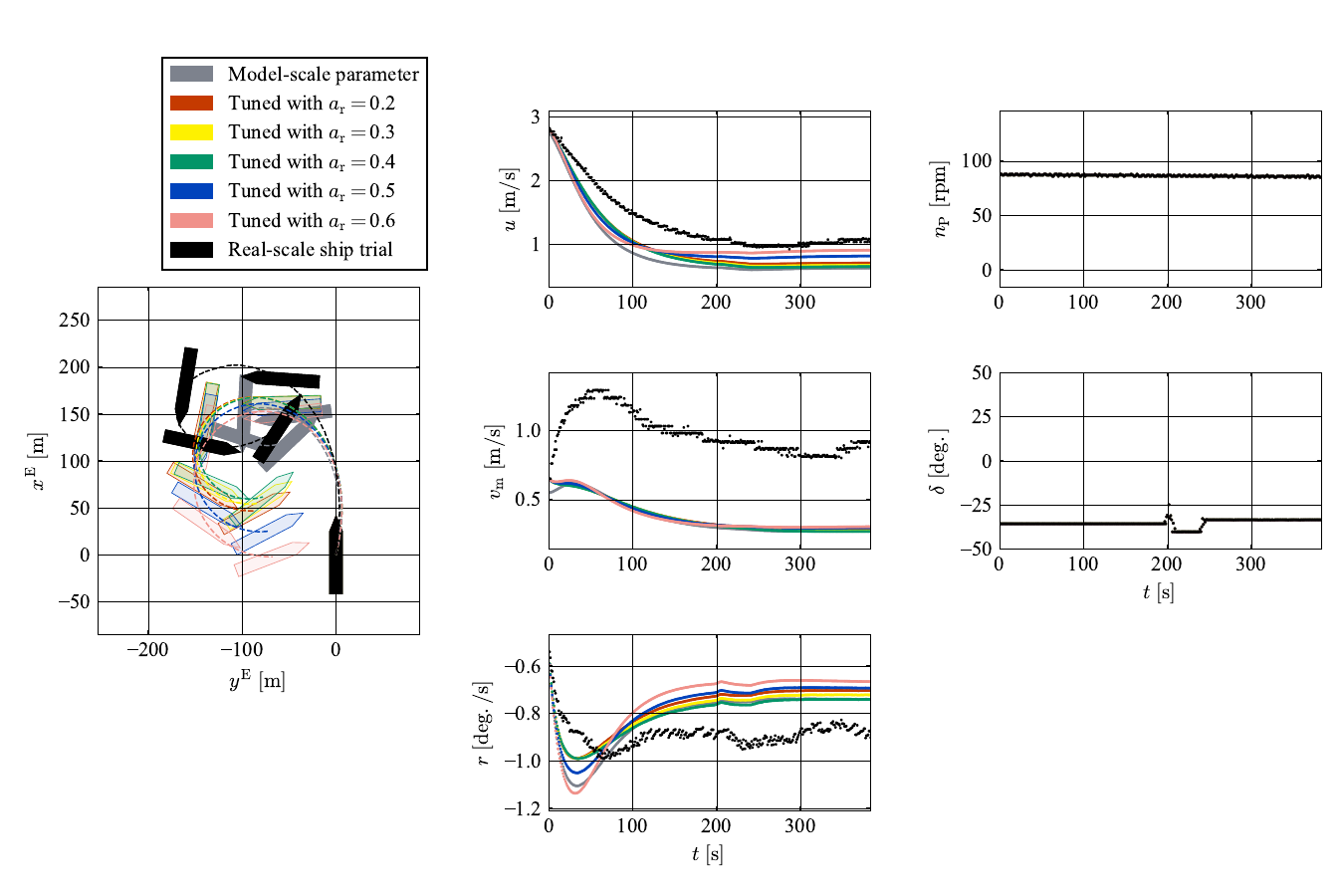}
            \caption{Comparison of seven time series data of turning test ($\delta = -35 ~ \mathrm{deg.}$) which is not utilized in parameter tuning: real-scale ship trial $D^{-35}$, simulation by the MMG model with untuned parameters, and simulation by the MMG model with tuned parameters ($a_{\mathrm{r}} = 0.2 \sim 0.6$).}
            \label{fig:compari_m35}
        \end{figure*}

\begin{figure*}[tb]
            \centering
            \includegraphics[width=1.0\hsize]{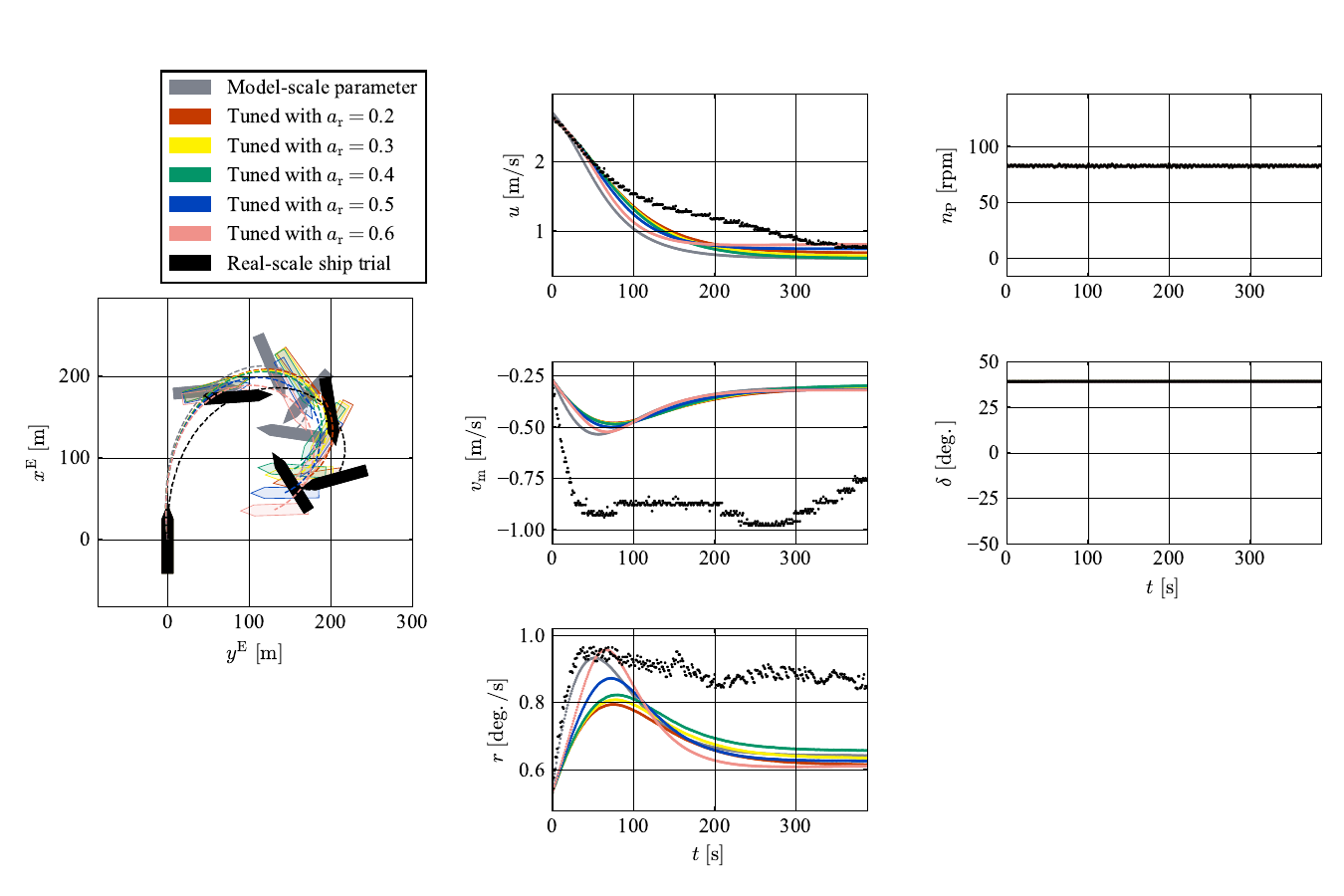}
            \caption{Comparison of seven time series data of turning test ($\delta = +40 ~ \mathrm{deg.}$) which is not utilized in parameter tuning: real-scale ship trial $D^{+40}$, simulation by the MMG model with untuned parameters, and simulation by the MMG model with tuned parameters ($a_{\mathrm{r}} = 0.2 \sim 0.6$).}
            \label{fig:compari_p40}
        \end{figure*}

        Finally, the time series data of the four real-scale ship trials included in $\mathcal{D}^{\mathrm{test}}$ were compared with the time series simulated based on the MMG model with the pre-determined and tuned parameter.
        The comparisons are shown in \cref{fig:compari_m10,fig:compari_p20,fig:compari_m35,fig:compari_p40}, respectively.
        All of the real-scale ship trials shown in \cref{fig:compari_m10,fig:compari_p20,fig:compari_m35,fig:compari_p40} were not used in the parameter tuning phase.
        By comparing these with the simulated time series using the MMG model with the tuned parameters, the estimation performance of the tuned parameter on the real-scale ship maneuvering motion can be evaluated.
        From trajectories in the left figure of each of \cref{fig:compari_m10,fig:compari_p20,fig:compari_m35,fig:compari_p40}, in all cases of $a_{\mathrm{r}} = 0.2 \sim 0.6$, the MMG models with the tuned parameter simulate time series closer to the maneuvering motion of the real-scale ship than the one with the pre-determined values.
        Comparing the best-performing case $a_{\mathrm{r}} = 0.5$ (blue-colored) and the worst-performing case $a_{\mathrm{r}} = 0.6$ (coral pink-colored), although the turning circle diameter differs by about $1.5 \Lpp$ for $D^{-10}$, not much difference in simulated maneuvering motion is observed for $D^{+20}$, $D^{-35}$, and $D^{+40}$.

\section{Discussion}

    \label{sec:discussion}

    From \cref{fig:compari_m10,fig:compari_p20,fig:compari_m35,fig:compari_p40}, it is confirmed that the parameters were tuned to simulate the turning motions with larger diameters than pre-determined parameters in all cases.
    Here, \cref{fig:explored_values_plot} shows that $a_{\mathrm{H}} > 0$ is tuned to be large and $x'_{\mathrm{H}} < 0$ is tuned to be small in all cases, suggesting that $N_{\mathrm{R}}$ is computed to be large.
    On the other hand, $w_{\mathrm{P}0}$ and $t_{\mathrm{R}}$ were tuned to be large so that the propeller inflow velocity $u_{\mathrm{P}}$ and the fraction of rudder $(1 - t_{\mathrm{R}}) F_{\mathrm{N}} \sin \delta$ are computed to be small, respectively.
    It is analyzed that these effects are dominant compared to the effect of largely estimated rudder moment $N_{\mathrm{R}}$.

    Among the parameter tunings conducted in this study, the parameter with the best performance on the test data was obtained with the exploration range defined by $a_{\mathrm{r}} = 0.5$.
    However, the proposed tuning method depends on the time series data set of the real-scale ship trial used for fine-tuning.
    When tuning the parameters of a maneuvering model, unexpected problems such as overfitting may occur depending on the time series data used for tuning, the target parameters to be tuned, and the exploration range of each parameter, as observed in \cref{fig:J_test_bar}.
    Therefore, for parameters explored under different conditions, it is not possible to determine which output is the best for the simulation of the real-scale ship maneuvering motion without at least checking the accuracy of simulated maneuvering motion for time series data other than $\mathcal{D}^{\mathrm{tune}}$, such as $\mathcal{D}^{\mathrm{test}}$ in this study.
    Thus, if there is a sufficient amount of time series data available for parameter tuning, it is desirable to prepare test data separately from the tuning data and compare the performance against validation data to determine the best parameter.

    However, the amount of available time series data of real-scale ship maneuvering motion is limited.
    Therefore, there may be cases where it is not possible to prepare validation data.
    In such cases, parameters tuned in a certain condition would have to be used in practice without performance validation.
    In this case, one should be careful not to set an unnecessarily wide exploration range, since parameters may be explored biased toward the tuning data.

\section{Concluding remarks}
    
    \label{sec:conclusion}

    An automatic fine-tuning method for all of the arbitrarily indicated target parameters of the MMG model was proposed.
    The proposed method tunes the parameter values which are previously determined based on the hydrodynamics, captive model tests, and CFD to the ones for the MMG model of the real-scale ship using the framework of SI.
    The previously determined parameter values are utilized to constrain the tuned parameter values to the realistic ranges.
    By directly referring to the time series data of real-scale ship maneuvering motion, the proposed method can steadily improve the performance of the MMG model with the tuned parameter in terms of the accuracy of the simulated maneuvering motion.
    The proposed fine-tuning method was applied to a container ship and validated with 12 target parameters which is highly influential in the MMG model.
    In all cases of parameter fine-tuning conducted with different widths of exploration range, better parameters were obtained compared to untuned parameters in terms of the accuracy of the simulated real-scale trajectories.

\printcredits


\section*{Acknowledgment}

    This study was conducted as a part of the Nippon Foundation to Support Fully Autonomous Ship Development Project ``MEGURI2040''.
    The authors are thankful to Mr. Takuya Taniguchi (Osaka University) for a helpful discussion.
    Moreover, this study was supported by a Grant-in-Aid for Scientific Research from the Japan Society for Promotion of Science (JSPS KAKENHI Grant \#22H01701) and by the Fundamental Research Developing Association for Shipbuilding and Offshore (REDAS23-5(18A)).

\end{document}